\documentclass[fleqn,usenatbib]{mnras}

\usepackage{newtxtext,newtxmath}

\usepackage[T1]{fontenc}
\usepackage{ae,aecompl}

\usepackage{float}
\usepackage{psfrag}
\usepackage{xfrac}
\usepackage{color}
\usepackage{multicol}

\usepackage{graphicx}	
\usepackage{amsmath}	
\usepackage{amssymb}	

\newcommand{\msun}{M_{\odot}}
\newcommand{\qh}{Q_{\rm thermal}}
\newcommand{\qgh}{Q_{\rm g,thermal}}
\newcommand{\HI}{{\sc H\,i}}
\newcommand{\HII}{{{\sc H\,ii}}}
\newcommand{\NII}{{{\sc N\,ii}}}
\newcommand{\OII}{{{\sc O\,ii}}}
\newcommand{\OIII}{{{\sc O\,iii}}}

\title[Conditions for Star Formation in Galaxy Discs]{SDSS-IV MaNGA: Constraints on the Conditions for Star Formation in Galaxy Discs}

\author[D. V. Stark et al.]{David V. Stark$^{1\dagger}$, Kevin A. Bundy$^{2}$, Matthew E. Orr$^{3}$, Philip F. Hopkins$^{3}$, \newauthor Kyle Westfall$^{2}$, Matthew Bershady$^{4}$, Cheng Li$^{5,6}$, Dmitry Bizyaev$^{7,8,9}$, \newauthor Karen L. Masters $^{10}$, Anne-Marie Weijmans $^{11}$, Ivan Lacerna$^{12,13,14}$, Daniel Thomas$^{10}$, \newauthor Niv Drory$^{15}$, Renbin Yan$^{16}$, and Kai Zhang$^{16}$
\\
$^{1}$ Kavli IPMU (WPI), UTIAS, The University of Tokyo, Kashiwa, Chiba 277-8583, Japan \\
$^{2}$ UCO/Lick Observatory, University of California, Santa Cruz, 1156 High St. Santa Cruz, CA 95064, USA\\	
$^{3}$TAPIR, 1200 E. California Blvd., California Institute of Technology, Pasadena, CA 91125, USA\\
$^{4}$University of Wisconsin-Madison, Department of Astronomy, 475 N. Charter Street, Madison, WI 53706-1582, USA\\
$^{5}$Physics Department and Tsinghua centre for Astrophysics, Tsinghua University, Beijing 100084, China\\
$^{6}$Shanghai Astronomical Observatory, Nandan Road 80, Shanghai 200030, China\\
$^{7}$Apache Point Observatory and New Mexico State University, PO Box 59, Sunspot, NM 88349-0059, USA\\
$^{8}$Sternberg Astronomical Institute, Moscow State University, 119992 Moscow, Russia\\
$^{9}$Special Astrophysical Observatory of the Russian AS, 369167, Nizhnij Arkhyz, Russia\\
$^{10}$Institute of Cosmology and Gravitation, University of Portsmouth, Dennis Sciama Building, Portsmouth PO1 3FX, UK\\
$^{11}$School of Physics and Astronomy, University of St Andrews, North Haugh, St Andrews KY16 9SS, UK\\
$^{12}$Instituto Milenio de Astrof\'isica, Av. Vicu\~na Mackenna 4860, Macul, Santiago, Chile\\
$^{13}$Instituto de Astrof{\'i}sica, Pontificia Universidad Cat{\'o}lica de Chile, Santiago, Chile\\
$^{14}$Astrophysical Research Consortium, Physics/Astronomy Building, Rm C319, 3910 15th Avenue NE, Seattle, WA 98195, USA\\
$^{15}$McDonald Observatory, The University of Texas at Austin, 2515 Speedway, Stop C1402, Austin, TX 78712, USA\\
$^{16}$Department of Physics and Astronomy, University of Kentucky, 505 Rose Street, Lexington, KY 40506-0055, USA\\
$^{{\dagger}}$david.stark@ipmu.jp\\
}

\date{Accepted 2017 October 30. Received 2017 October 9; in original form 2017 July 27}

\pubyear{2017}

\begin{document}
\label{firstpage}
\pagerange{\pageref{firstpage}--\pageref{lastpage}}
\maketitle

\begin{abstract}
Regions of disc galaxies with widespread star formation tend to be both gravitationally unstable and self-shielded against ionizing radiation, whereas extended outer discs with little or no star formation tend to be stable and unshielded on average.  We explore what drives the transition between these two regimes, specifically whether discs first meet the conditions for self-shielding (parameterized by dust optical depth, $\tau$) or gravitational instability (parameterized by a modified version of Toomre's instability parameters, $\qh$, which quantifies the stability of a gas disc that is thermally supported at $T=10^4$ K). We first introduce a new metric formed by the product of these quantities, $\qh\tau$, which indicates whether the conditions for disk instability or self-shielding are easier to meet in a given region of a galaxy, and we discuss how $\qh\tau$ can be constrained even in the absence of direct gas information. We then analyse a sample of 13 galaxies with resolved gas measurements and find that on average galaxies will reach the threshold for disk instabilities ($\qh<1$) before reaching the threshold for self-shielding ($\tau>1$). Using integral field spectroscopic observations of a sample of 236 galaxies from the MaNGA survey, we find that the value of $\qh\tau$ in star-forming discs is consistent with similar behavior. These results support a scenario where disc fragmentation and collapse occurs before self-shielding, suggesting that gravitational instabilities are the primary condition for widespread star formation in galaxy discs. Our results support similar conclusions based on recent galaxy simulations.
\end{abstract}




\section{Introduction}
\label{sec:intro}
Star formation can only proceed within galaxies when certain conditions are met. For decades, several physical models for star formation have been proposed and debated, typically relating the star formation rate (SFR) to specific properties of the interstellar medium (ISM) such as gas surface density, metallicity, dynamics, or some combination thereof.  The most popular star formation relation was introduced by \citet{Schmidt59} who argued that the local surface density of star formation, $\Sigma_{\rm SFR}$, is related to the local surface density of gas, $\Sigma_{\rm g}$, by the power law $\Sigma_{\rm SFR} \propto \Sigma_{\rm g}^N$.  \citet{Kennicutt98} showed that globally averaged values of $\Sigma_{\rm g}$ and $\Sigma_{\rm SFR}$ follow such a relation over several orders of magnitude.  A similar correlation is found on sub-kpc scales \citep{Bigiel08} with an additional dependence on the local dust-to-gas ratio \citep{Leroy13}.  At gas densities below $\Sigma_{\rm g}\sim10\,{\rm M_{\odot}\,pc^{-2}}$ the power law relation breaks down, but it can be recovered if one isolates only the denser molecular gas in regions which are on-average low density \citep{Schruba11}, consistent with a picture where $\Sigma_{\rm SFR}$ is most directly correlated with the denser molecular gas surface density, $\Sigma_{\rm H_2}$.  

Under the assumption that $\Sigma_{\rm SFR}\propto\Sigma_{\rm H_2}$, several authors have formulated star formation laws based around conditions of the ISM that govern its molecular fraction.  \citet{Elmegreen89} introduced a relationship between molecular fraction and hydrostatic midplane pressure (see also \citealt{Elmegreen94}, \citealt{Wong02}, \citealt{Blitz04}, \citealt{Blitz06}).  Alternative models assume H$_2$ formation is regulated by local gas density and metallicity (or dust-to-gas ratio, assumed to follow a 1:1 relation with metallicity), the combination of which determine the ability for gas to shield itself from background ionizing radiation, at which point it can cool, condense, and form stars \citep{Schaye04,Krumholz09,Krumholz12}. A common theme of these prescriptions is that star formation is regulated by the local conditions of the ISM, and in principle the relevant physics should not vary from galaxy to galaxy. 

However, several studies have suggested that the larger-scale dynamical properties of galaxies play a major role in regulating star formation.  In such models, the potential for dense clouds to form is set by the competition between the self-gravity of the disc and some combination of gas dispersion, Coriolis forces, cloud collision rate, and shear
\citep{Safronov60,Toomre64,Jog84,Romeo92,Wang94,Hunter98,Tan00, Rafikov01,Elmegreen11}. Prescriptions which relate $\Sigma_{\rm SFR}$ to dynamical properties (or dynamical timescales) are able to fit both regular star-forming galaxies and starbursts on a single relation, which has been a struggle for laws which relate star formation and gas density alone \citep{Daddi10,Genzel10,Utreras16}. Additional support for the relevance of dynamical properties comes from observations of ``star formation thresholds," rapid drops in the $\Sigma_{\rm SFR}$ at the presumed radius where discs stabilize \citep{Kennicutt89,Martin01} according to the $Q$ parameter of \citet{Toomre64}. However, these interpretations have been challenged by commonly observed (although typically weak) star formation at radii beyond these ``thresholds".  In particular, studies that trace star formation using UV continuum find much less extreme radial declines in star formation compared to studies which use H$\alpha$ emission \citep{Ferguson98,Ryan-Weber04,GildePaz05,Boissier07,Thilker07,Werk10,Christlein10,Hunter10,Lemonias11,Moffett12}. One physical explanation for the observed difference between H$\alpha$ and UV profiles is that SFR has dropped low enough in outer discs that there are simply too few massive, short lived O stars to generate H$\alpha$ emission, whereas the UV emission is more visible because it is sensitive to slightly less massive, longer lived B stars \citep{Boissier07}. None the less, studies have found that star formation efficiency, or SFE (here defined as ${\rm SFE=\sfrac{\Sigma_{\rm SFR}}{\Sigma_{\rm g}}}$) is significantly lower (by a factor of $\sim$10) in extended outer discs compared to inner discs \citep{Kennicutt89,Zasov94,Bigiel10b}.  It is possible that although star formation can occur at large radii in the presence of localized overdensities -- due to e.g., tidal interactions, spiral density waves, and/or cold accretion \citep{Thilker07,Bush08,Roskar10} -- outer discs are on-average dynamically stable against fragmentation, explaining their extremely low SFE. 

Many of the proposed models for star formation are at least partially valid because they relate the conditions of the ISM to the presence and/or rate of star formation.  Nevertheless, it is still debated which physical prescription best describes the underlying conditions that eventually lead to widespread star formation in galaxies. As an alternative way of posing this question, let us consider radial annuli within a galaxy. At some large radius the gas in an annulus will be dynamically stable (likely supported by thermal gas pressure at $T \gtrsim 10^4\,K$) and not self-shielded against the background UV field.  Presumably, this annulus will have no star formation, or at least extremely low efficiency of star formation (as mentioned above, density fluctuations may drive localized star formation, but in an annulus-averaged sense, the disc is not conducive to star formation).  Conversely, at some smaller radius, the disc will be unable to resist fragmentation, will be shielded against background radiation, and will be forming stars.  {\it What defines this transition?}  As we move from large to small radius, do annuli first become unstable and fragment,  only after which the gas self-shields?  Or, does the gas reach its self-shielding threshold first, only after which instabilities and fragmentation can occur? 

Multiple studies have attempted to address this question, with differing results.  Work by \citet{Schaye04} and \citet{Krumholz09} suggest that instability occurs only after a cold ISM phase develops which lowers the gas velocity dispersion. In these models, star formation is tied to the dust optical depth, $\tau$, which determines the ability of gas to self-shield.  In contrast, \citet{Orr17} examine the FIRE simulations \citep{Hopkins14} and find that star formation is prevalent throughout regions which are not self-shielded on average.  Instead, the onset of vigorous star formation ($\Sigma_{\rm SFR}>10^{-3}\,{\rm M_{\odot}\,yr^{-1}\,kpc^{-2}}$) occurs very close to where galaxies cross the threshold where discs can no longer support themselves through thermal gas pressure ($Q_{\rm thermal} \lesssim 1$), which occurs well before they cross the threshold for self-shielding. In fact, additional tests where \citet{Orr17} remove the ability of gas to self-shield and cool below $10^4\,K$ yield distributions of star formation that are very similar to the full physics runs.  Meanwhile, tests where {\it only} self-shielded gas can form stars yield almost no star formation beyond $\sim1/3R_e$ (where $R_e$ is the half-light radius), in significant disagreement with the full physics runs. The discrepancies between different studies highlight the need for observational studies to constrain which of these theoretical pictures is correct.

We use two independent data sets to observationally constrain the link between star formation, disc stability ($Q_{\rm thermal}$), and dust optical depth ($\tau$). The first data set, drawn from the compilation of \citet{Leroy08}, is composed of a small sample of nearby galaxies with high quality measurements of gas content. The second sample is drawn from the significantly larger and more representative MaNGA survey \citep{Bundy15}, but lacks estimates of gas surface density needed to estimate $\tau$ and $\qh$ directly.  However, we describe how the product of these two parameters, $\qh \tau$, can be used to distinguish whether the conditions for self-shielding or gravitational instability will be met first, and can be constrained without direct gas measurements. Thus, we address the science questions of this paper from two different angles with data sets that each have their respective strengths and weaknesses.  

In Section~\ref{sec:methods} we describe the parameters used to quantify the conditions of the ISM and how these parameters are used to test different physical models of the onset of star formation in galaxies. In Section~\ref{sec:data}, we describe the data sets used in this study. Our results are presented in Section~\ref{sec:results}, and in Section~\ref{sec:discussion} we give a more detailed discussion of important systematic errors and the implications of our findings.  Our conclusions are presented in Section~\ref{sec:conclusions}.

\section{Methodology} 
\label{sec:methods}

\subsection{$\tau$ and $\qh$}
\label{sec:qtau}
We consider two key parameters to characterize the average conditions of the ISM. The first parameter is $\tau$, a proxy for dust optical depth:
\begin{equation}
\label{eq:tau}
\tau=\frac{\Sigma_{\rm g}Z'}{\left(\Sigma_{\rm g} Z'\right)_{\rm ss}}
\end{equation}
where $\Sigma_{\rm g}$ is the gas surface density, $Z'$ is the gas-phase metallicity relative to solar, and $\left(\Sigma_{\rm g} Z'\right)_{\rm ss}$ represents the dust surface density above which the ISM is self-shielded and ${\rm H_2}$ dominated (${\Sigma_{\rm H_2}}/{\Sigma_{\rm g}} > 0.5$) in the presence of an isotropic background ionizing field. The specific value of $\left(\Sigma_{\rm g} Z'\right)_{\rm ss}$ depends on metallicity and is calculated using Eq.~45 of \citet{Krumholz09}.  For solar metallicity gas, \mbox{$\left(\Sigma_{\rm g} Z'\right)_{\rm ss}=27\,\msun\,{\rm pc^{-2}}$} and varies from \mbox{$18-36 \, \msun \,{\rm pc^{-2}}$} between \mbox{$Z'=0.1-3$}.  Previous implementations of the \citet{Krumholz09} model have employed a clumping factor to account for unresolved individual clouds.  However, as we are interested in the {\it average} ability an annulus to self-shield, we include no such factor.

The second parameter used to characterize the ISM is a modified version of the \citet{Toomre64} $Q$ parameter and indicates the ability of the disc to support itself against fragmentation through Coriolis forces and {\it thermal} gas pressure:
\begin{equation}
\label{eq:q}
\qgh=\frac{\kappa c_s}{\pi G \Sigma_{\rm g}}
\end{equation}
where $\kappa$ is the epicyclic frequency and $c_s$ is the characteristic sound speed of $10^4\,{\rm K}$ gas ($13\,{\rm km\,s^{-1}}$) 
  
$\qgh$ is not the same as $Q$ commonly measured in cold gas. $\qgh$ quantifies the ability of the disc to stabilize itself explicitly via {\it thermal} gas pressure assuming all the gas has $T=10^4\,{\rm K}$, which characterizes the typical conditions in the non-star forming outer gas disc. We expect that regions with $\qgh<1$ will fragment and form stars, which in reality can generate other forms of support (e.g., turbulence) through gravitational instabilities and feedback that drive the {\it observed} $Q$ back towards $\sim$1 even though $\qh < 1$. However, the fact that such regions are experiencing non-thermal means of support, which either lead to star formation or result from it, suggests that they have entered a ``star-forming" regime.  Therefore, $\qgh$ is a conservative indicator of the ability of a gaseous annulus to resist fragmentation, while also capturing the relevant conditions of the outer gas disc.  
  
The presence of a significant stellar component will contribute to the overall stability of the disc and must be taken into account.  The stellar disc stability is defined as
\begin{equation}
Q_*=\frac{\kappa\sigma_*}{\pi G \Sigma_*}
\end{equation}
where $\sigma_*$ is the radial stellar velocity dispersion and $\Sigma_*$ is the disc stellar surface density. The overall stability of the two-component disc, $\qh$ can then be approximated as \citep{Wang94}\footnote{We note that more accurate approximations of a multi-component disc have been presented by \citet{Romeo11} and \citet{Romeo13}. These approximations raise $Q_{\rm thermal}$, but not enough to change the conclusions of this paper.}
\begin{equation}
\qh=\left(\frac{1}{\qgh}+\frac{1}{Q_*}\right)^{-1}
\end{equation}
    
  
Although we have chosen $\qh$ as our indicator of disc stability, it is not the only proposed means to describe the threshold for gravitational instability.  \citet{Hunter98} proposed a scenario in which the formation of dense clouds is regulated by competition with shear.  Their formalism primarily affects inner regions of galaxies with rising rotation curves, whereas our study will be focused on larger radii where rotation curves are typically flat, in which case the shear-regulated threshold reduces to approximately the same form as $\qh$.  

\subsection{The Onset of Widespread Star Formation and $\qh\tau$}
\label{sec:qtau_model}

\begin{figure}
	\includegraphics[width=\columnwidth]{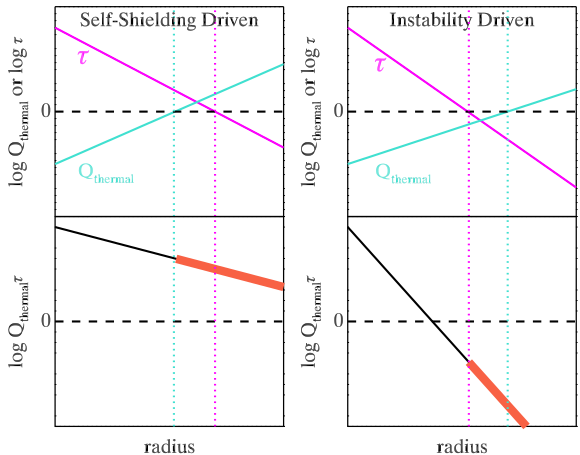}
	\caption{Schematic models illustrating the radial behavior of $\qh$ and $\tau$ (top panels), and $\qh\tau$ (bottom panels) under two different scenarios to describe the onset of widespread star formation in galaxy discs. The horizontal dashed line indicates the threshold for star formation, and the vertical dotted lines indicate where each parameter crosses this threshold.  The thick red line in the bottom panels highlight the region in the outer disc where $\qh\tau$ can be used to discriminate the two scenarios. Even in regions where neither threshold has been crossed (e.g., at very large radius in both models), $\qh\tau$ still indicates which threshold will be crossed first if the gas density is increased.}
\label{fig:tau_q_model}
\end{figure}

We consider two basic scenarios that describe the transition between non/weakly star-forming outer discs to vigorously star-forming inner regions of galaxies.  This simple picture assumes that at some large radius, the gas disc is gravitationally stable from thermal pressure ($\qh>1$) and unshielded ($\tau<1$), while the centre is gravitationally unstable ($\qh<1$) and self-shielded ($\tau>1$). The two scenarios differ in their assumptions about whether self-shielding or disc instabilities determine the onset of widespread star formation in galaxy discs as we move from large to small radius, which is essentially a question of which threshold for star formation is met first.  These two scenarios are illustrated in the top row of Fig.~\ref{fig:tau_q_model}.
\begin{enumerate}
\item {\it Self shielding-driven}: disc instabilities and fragmentation occur {\it after} self-shielding (e.g., \citealt{Schaye04}, \citealt{Krumholz09}). In this scenario, a galaxy will cross the threshold for self-shielding {\it before} it crosses the threshold for disc instabilities. Therefore, as we move from large to small radius, we will encounter annuli that are on-average self-shielded ($\tau > 1$) but still gravitationally unstable ($\qh>1$).
\item {\it Fragmentation-driven}: disc instabilities occur {\it before} self-shielding (e.g., \citealt{Orr17}). In this scenario, the threshold for gravitational instability from thermal support is crossed before the threshold for self-shielding. Therefore, as we move from large to small radius, we will encounter annuli with that are on-average gravitationally unstable ($\qh<1$) but not yet self-shielded ($\tau<1$). 
\end{enumerate} 

Note that under the second scenario where $\qh<1$ and $\tau<1$, we are not implying that any {\it localized} star forming regions are unshielded. Instead, they are almost definitely self-shielded, but since self-shielding occurs after disc fragmentation, the point at which the disc {\it on-average} has $\tau>1$ lags behind the point at which the disc {\it on-average} has $\qh<1$.

Both $\tau$ and $\qh$ depend on $\Sigma_{\rm g}$, which is often unknown.  However, we argue that the product, $\qh\tau$ can provide useful constraints on whether fragmentation or disc opacity are the underlying driver of widespread star formation in galaxy discs, and this product can be constrained even in the absence of direct gas estimates.  

For sufficiently gas-dominated regions of galaxies where $\Sigma_{\rm g} \gg \Sigma_*$, we can approximate $\qh\sim\qgh$. In the product, $\qh\tau$, the $\Sigma_{\rm g}$ terms cancel yielding the following functional form 
\begin{equation} \label{eq:qtau_approx}
\qh\tau \leq \frac{\kappa c_s Z'}{\pi G \left(\Sigma_{\rm g} Z'\right)_{\rm ss}}
\end{equation}
The above formula is written as an upper limit because the inclusion of any stellar component will typically lead to a lower value for $\qh\tau$.  

Although we would ideally compare independent measures of $\qh$ and $\tau$ against one another, we can still place constraints on $\qh\tau$ without such information. To illustrate the utility of $\qh\tau$ further, we revisit the two basic scenarios shown in Fig.~\ref{fig:tau_q_model}:
\begin{enumerate}
\item {\it Self shielding-driven}:This scenario is characterized by the presence of annuli which are on-average self-shielded ($\tau>1$) but not gravitationally unstable ($\qh>1$).  These annuli will have $\qh\tau>1$.
\item {\it Fragmentation-driven}: This scenario is characterized by the presence of annuli which are on-average gravitationally unstable ($\qh>1$) but not self-shielded ($\tau<1$). These annuli will have $\qh\tau<1$.
\end{enumerate} 
The behavior of $\qh\tau$ is illustrated in the bottom panels of Fig.~\ref{fig:tau_q_model}. The key difference between these scenarios is whether $\qh\tau$ is greater than or less than unity in the outer discs of galaxies. Formally, $\qh\tau$ measures whether $\qh$ or $\tau$ is closest to its relevant threshold, but does not imply either threshold is currently met. In the case where neither threshold is satisfied (e.g., at the largest radii in Fig.~\ref{fig:tau_q_model}), $\qh\tau$ indicates which threshold would be {\it easier} to reach if the gas surface density were larger. In this sense, $\qh\tau$ allows us to ``roll back the clock'' in annuli where star formation is occurring but neither threshold is currently satisfied; we can infer which threshold is more likely to have been reached first to initiate star formation in the past.
 
We must be cautious when interpreting $\qh\tau$ at small radii. $\qh\tau$ is informative at and beyond the region where one threshold has been reached, but the other has not. For example, in Fig.~\ref{fig:tau_q_model} $\qh\tau>1$ for both the self-shielding and instability-driven models at small radius.  For this reason, we will focus our analysis on the value of $\qh\tau$ at large radii (typically $>1.5R_e$). 
 
Throughout this work, we define \mbox{$\Sigma_{\rm SFR}>10^{-3}\,M_{\odot}\,{\rm yr^{-1}\,kpc^{-2}}$} as the boundary between regions with widespread star formation and those with weak/no star formation. This specific value has little physical significance. Although it roughly corresponds to the typical value where $\Sigma_{\rm SFR}$ profiles tend to more rapidly decline (star formation ``thresholds''), such behavior may simply reflect the failure of H$\alpha$ emission as a star formation tracer at low surface brightness. We largely highlight $10^{-3}\,M_{\odot}\,{\rm yr^{-1}\,kpc^{-2}}$ to facilitate comparison with \cite{Orr17}, who use the same value to separate star-forming and non-star forming annuli. The key results of this paper are not explicitly tied to this value of $\Sigma_{\rm SFR}$.

Choosing $\Sigma_{\rm SFR}=10^{-3}\,M_{\odot}\,{\rm yr^{-1}\,kpc^{-2}}$ as the definition of ``star forming'' annuli also essentially guarantees that star formation is present in annuli which are currently not self-shielded on average.  Our definition of self-shielded corresponds to $\Sigma_{\rm H_2}/\Sigma_{\rm HI} \sim 1$, which is expected in gas densities of ${\sim}27\,M_{\odot}\,{\rm pc^{-2}}$ \citep{Krumholz09}, and these gas densities typically correspond to $\Sigma_{\rm SFR}\sim10^{-2}\,M_{\odot}\,{\rm yr^{-1}\,kpc^{-2}}$ \citep{Bigiel08}. This fact does not immediately rule out the ``self-shielding driven'' star formation scenario. Instead, the important test being conducted in this study is whether disc instabilities occur before or after self-shielding. It is more important to consider the conditions of the ISM at the start of star formation rather than at the present, and as discussed above, $\qh\tau$ allows us to consider which threshold would have been easier to reach first.

\section{Data and Derived Quantities}
\label{sec:data}

We employ two independent data sets for our analysis. The first, drawn from \citet[hereafter L08]{Leroy08} contains direct measurements of gas surface density for a small subset of nearby galaxies.  The second, drawn from the MaNGA survey, lacks direct estimates of gas surface density but is significantly larger and more representative of the galaxy population. In the following section, we discuss these data sets and derived quantities.

\subsection{The \citet[L08]{Leroy08} sample}
To compose a sample of galaxies with resolved gas data, we use the compilation from \citet{Leroy08} who combine data from the SINGS \citep{Kennicutt03}, THINGS \citep{Walter08}, HERACLES \citep{Leroy09}, BIMA SONG \citep{Helfer03}, and GALEX NGS \citep{GildePaz07} surveys to create radial surface density profiles of \HI, H$_2$, stellar mass, and SFR with sub-kpc radial spacing for 23 nearby galaxies.  We refer the reader to \citet{Leroy08} for more specific details about the calculation of these profiles.

\subsubsection{Kinematics}

Rotation velocities, $v_{\rm rot}$, were derived from THINGS velocity fields.  \citet{Leroy08} provide fits to these data of the form $v_{\rm rot}=v_{\rm flat}\left[1-e^{-r/l_{\rm flat}}\right]$.  We use the analytical fits to estimate $\kappa$ as a function of radius.

Stellar velocity dispersions, $\sigma_*$ are not readily available for the L08 sample, so we estimate them indirectly assuming
\begin{equation}
\sigma_*=\sqrt{\pi G \left(\Sigma_{\rm g} + \Sigma_*\right)z_0}
\label{eq:sigma_s}
\end{equation}
where $z_0$ is the disc vertical scale parameter for a disc whose density decreases with height $z$ above the midplane as ${\rm sech}^2(z/z_0)$ \citep{VanDerKruit81}.  We estimate $z_0$ using the median ratio of disc scale length ($h$) to scale height in the SDSS $r$-band, $h/z_0=3.4$ \citep{Bizyaev14}.  The disc scale lengths are taken from the exponential fits to stellar discs given by L08.  Note that our estimates of $\sigma_*$ may be poor at small radius where the bulge component becomes important.

\subsubsection{Gas Phase Metallicity}

Gas-phase metallicities are taken from \citet{Moustakas10} who compile metallicities of individual {\HII} regions in galaxies from the SINGS sample.  This compilation provides two different estimates of metallicity using the calibrations of \citet{Kobulnicky04} (KK04) and \citet{Pilyugin05} (PT05), both of which employ the $R_{23}$ parameter \citep{Pagel79}.  Due to the notoriously large systematic errors between different strong-line calibrations, we must be cautious when choosing our metallicity estimate and drawing conclusions from it.  We opt to use the KK04 metallicities because of their relatively good agreement with the N2O2 calibration we employ for the MaNGA sample (see Section~\ref{sec:manga_metal}); the two calibrations yield mass-metallicity relationships with similar slopes and systematic offsets no larger than 0.1 dex \citep{Kewley08}. The impact of systematic errors in metallicity estimates is discussed further in Section~\ref{sec:results}.

Since the \citet{Moustakas10} compilation provides metallicities for individual {\HII} regions, it does not always evenly sample the metallicity profile at all radii.  Therefore, to estimate gas-phase metallicity as a function of radius, we fit linear functions to the {\HII} region metallicities ($12+\log {\rm O/H}$) as a function of radius for each galaxy.  We require at least 5 data points for our linear fits.

\subsubsection{Sample Selection}
We limit our sample to galaxies with stellar masses above $M_*=10^9\,M_{\odot}$ to match the approximate minimum stellar mass of MaNGA.  After removing any additional galaxies with insufficient metallicity information, we are left with 13 galaxies in our final sample.   

\subsection{MaNGA Data}

Our second data set comes from the SDSS-IV MaNGA survey \citep{Bundy15,Drory15,Law15,Yan16b,Yan16a,Law16,SDSS16,Blanton17}, an integral field unit (IFU)
survey of 10,000 $z{\sim}0$ galaxies with $M_*>10^9\,\msun$. This survey uses the SDSS 2.5m telescope \citep{Gunn06} and BOSS
spectrographs \citep{Smee13}, with a wavelength coverage of \mbox{3500--10000 {\AA}}, spectral resolution $R\sim2000$
(instrumental resolution ${\sim}60\,{\rm km\,s^{-1}}$), and an effective spatial resolution of 2.5$\arcsec$ (FWHM) after combining dithered observations.

\subsubsection{Kinematics}
\label{sec:kinematics}

Assuming a parametric form for rotation curves, tilted thin-disc models from \citet{Andersen13}, which determines the best-fitting kinematic geometry
(position angle, PA, and inclination, $i$), are fit to stellar and ionized gas (H$\alpha$) kinematic fields from the MaNGA data
analysis pipeline.  One dimensional profiles sampled every 2.5$\arcsec$ are extracted using the data within
$\pm30^{\circ}$ of the major axis. These profiles are fit with a model of the form $v_{\rm rot}(R)=v_{\rm
  flat}\tanh{\left(R/h_{\rm rot}\right)}$, where $v_{\rm flat}$ and $h_{\rm rot}$ are free parameters, in order to estimate
$\kappa$. Although this formula is different from that employed by \citet{Leroy08}, the two have a very similar shape and we do not expect this difference to impact our results.

Our analysis is limited to disc galaxies that are ``kinematically regular,'' defined by similar kinematic geometries for the
stars and gas and disc inclinations consistent with photometric estimates based on the NASA Sloan Atlas
(NSA)\footnote{http://www.nsatlas.org/}.  The kinematic modeling process is further described in K. Westfall et al. (in prep). 

Although MaNGA technically measures stellar velocity dispersions, the expected values are $\sim30\,{\rm km\,s^{-1}}$ in the outer discs of galaxies  \citep[e.g.][]{Bottema93,Shapiro03,Martinsson13}, well below the MaNGA's velocity resolution of $\sim60\,{\rm km\,s^{-1}}$.  Therefore, we again indirectly estimate $\sigma_*$ using Eq.~\ref{eq:sigma_s}. Disc scale lengths are taken from the bulge-disc decompositions of \citet{Simard11}.  



\subsubsection{Gas-Phase Metallicities}
\label{sec:manga_metal}

Two different strong-line calibrations are used to estimate gas-phase metallicity. The first, N2O2, uses the 
[\NII]$\lambda6584$/[\OII]$\lambda3727$ flux ratio and is based on photo-ionization models \citep{Kewley02}. It is relatively
insensitive to the ionization parameter and diffuse interstellar gas \citep{Zhang17}. The second method,
O3N2, uses the [\OIII]$\lambda5007$/H$\beta$ and [\NII]/H$\alpha$ flux ratios \citep{Marino13}, and while more sensitive to ionization
and diffuse interstellar gas, it is calibrated directly from observations using electron temperature (i.e., the ``direct" method). The N2O2 and O3N2 methods differ on their merits but helpfully bracket the
metallicity range of ${\sim}0.4$ dex spanned by available calibrations. We assume a solar metallicity of $12+\log{({\rm
    O/H})_{\odot}}=8.69$ \citep{Asplund09}.

All emission line fluxes are corrected for foreground extinction using \citet{Schlegel98}. Internal extinctions are estimated from
the Balmer decrement assuming \mbox{${\rm H}\alpha/{\rm H}\beta=2.86$} \citep{Osterbrock06}. We limit our analysis to regions
where the H$\alpha$ and H$\beta$ flux S/N is larger than 3. All corrections use the extinction curve 
from \citet{Fitzpatrick99} with $R_V$=3.1.

Radial metallicity profiles are extracted for each galaxy using physical radii determined from the kinematic modeling (see
Section \ref{sec:kinematics}).  Radial bins are spaced by $2.5\arcsec$ along the major axis. The final metallicity in each bin is the median of all spaxels in
that bin. We only use spaxels with [\OIII]/H$\beta$ and [\NII]/H$\alpha$ in the {\HII} region of the
Baldwin-Phillips-Terlevich (BPT) diagram \citep{Kewley06b}. We ignore any spaxels flagged as unreliable by the MaNGA data
reduction and analysis pipelines (\citealt{Law16}, K. Westfall et al. in prep) and only include annuli with at least five usable spaxels.

\subsubsection{Stellar Surface Densities}
Estimates of $\Sigma_*$ come from the \texttt{Pipe3D} analysis pipeline which uses a modified version of \texttt{FIT3D}, a fitting tool for analyzing the properties of stellar populations and ionized gas with moderate resolution optical spectra of galaxies, where linear combinations of SSPs are fit to each spaxel to determine $\Sigma_*$.  We refer the reader to \citep{Sanchez16a} and \citep{Sanchez16b} for further details about the fitting procedure. Radial profiles of $\Sigma_*$ are determined using the average of all spaxels in annuli identical to those used to calculate metallicity profiles.

\subsubsection{Sample Selection}
\label{sec:sample}
Our sample is drawn from data acquired during the first year of normal MaNGA operations (1368 galaxies), roughly equivalent to the SDSS-IV DR13\footnote{http://www.sdss.org/dr13} sample.  In addition to selecting only ``kinematically regular" galaxies (Section~\ref{sec:kinematics}), we only include galaxies with \mbox{$40^{\circ}<i<70^{\circ}$} to further ensure both reliable kinematics and that our analysis focuses on disc material rather than extraplanar gas. We remove any galaxies where the measured rotation curves have not flattened by the outermost radius so that we can reliably extrapolate to estimate $v_{\rm rot}$ at large radii along minor axes.  We reject galaxies with ${\rm NUV}-r>4.5$ to avoid extremely gas-poor systems (gas-to-stellar mass ratio $<0.05$; \citealt{Catinella13}), where the NUV and $r$-band magnitudes are taken from the NSA.  Lastly, we remove any galaxies which are obvious mergers. Our final sample is composed of 236 galaxies with stellar masses of ${\sim}10^{9-11.2}\,\msun$. 

\begin{figure}
\includegraphics[width=\columnwidth]{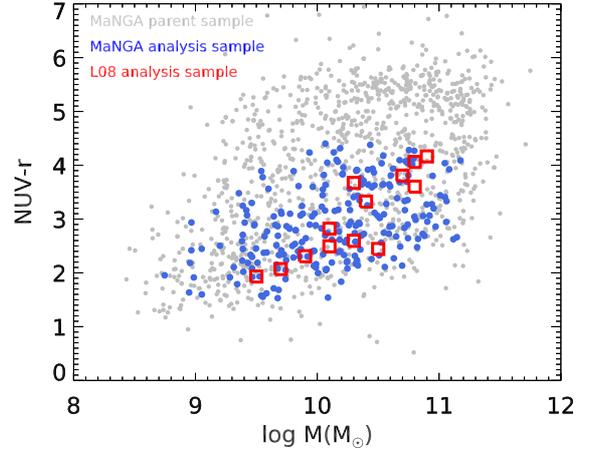}
\caption{${\rm NUV}-r$ vs. $M_*$ distribution for the parent MaNGA sample (gray), the subset of 236 MaNGA galaxies incorporated into our analysis (blue), and the 13 galaxies from L08 (red squares).}
\label{fig:sample}
\end{figure}

Fig.~\ref{fig:sample} shows our final sample in ${\rm NUV}-r$ vs. $M_*$ space in the context of the full MaNGA sample. We also show the position of our sample drawn from \citet{Leroy08} using photometry taken from \citet{Munoz-Mateos09} and \cite{Brown14}.  In three cases SDSS $r$-band magnitudes were not measured directly, in which case we roughly approximated them by interpolating the optical-IR SEDs.

\section{Results} \label{sec:results}

With our two independent samples, we investigate the behavior of $\qh$, $\tau$, and $\qh\tau$ in galaxy discs in order to test the self-shielding driven and fragmentation driven star formation scenarios described in Section~\ref{sec:qtau_model}.

\subsection{The L08 Sample}

\begin{figure*}
\includegraphics[width=2\columnwidth]{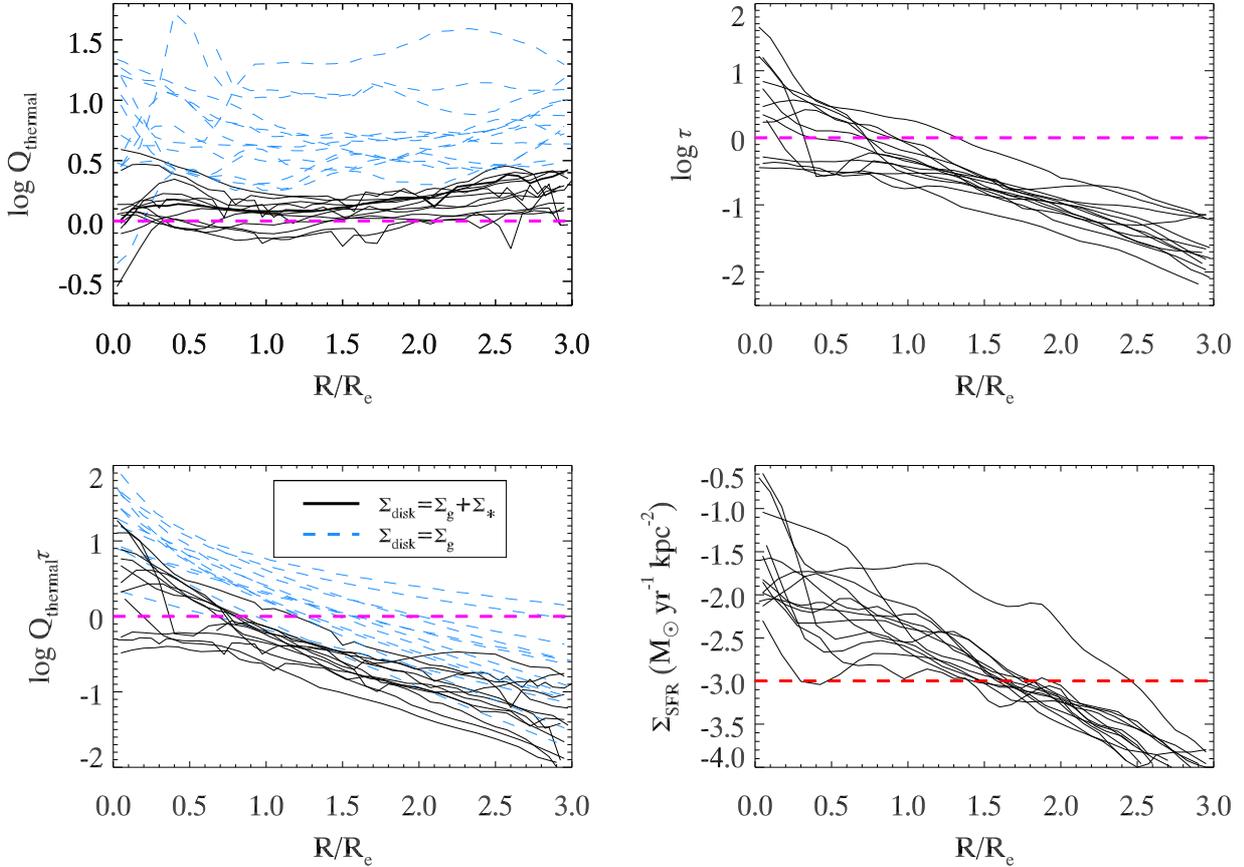}
\caption{Radial profiles of $\qh$, $\tau$, $\qh\tau$, and $\Sigma_{\rm SFR}$ for our sample of 13 galaxies with resolved gas data from L08.  Where applicable, the blue dashed lines show the estimates assuming $\Sigma_*$ is negligible (Eq.~\ref{eq:qtau_approx}).  The radii taken from \citet{Leroy08} are normalized by the optical disc radius, $R_{25}$.  We convert the radii to units of $R_e$ for easier comparison to our MaNGA sample by assuming $R_{25}/R_e=2.5$ (based on radii tabulated in \citealt{Jansen00b}). These data show that discs remain at least borderline unstable while their ability to self-shield drops off much more rapidly. Consistent with the behavior of $\tau$ and $\qh$,  $\qh\tau < 1$ over much of the outer disc.  }
\label{fig:qtau_wgas}
\end{figure*}

First, we examine the behavior of the L08 sample with direct measurements of gas content. Radial profiles of $\qh$, $\tau$, $\qh\tau$, and $\Sigma_{\rm SFR}$ are shown in Fig.~\ref{fig:qtau_wgas}.  We also separately show $\qh$ and $\qh\tau$ for just the gas components, giving an indication of how ignoring the stellar component can overestimate these quantities. Note that although we have plotted $\qh\tau$ as a function of radius, our primary interest is the value of $\qh\tau$ in the outer discs of galaxies.

While there is a mild trend of decreasing $\qh$ with decreasing radius, generally $\qh\sim1$ over all radii, signaling that the discs are neither drastically stable or unstable.  Notably, the gas discs alone would be stable through thermal support alone; it is the addition of the stellar component that weakens the overall stability of the disc.  The ability of the disc to self shield drops off very rapidly with increasing radius, and in some cases never reaches $\tau>1$.  Significant star formation is present in annuli which are not, on average, self-shielded (as previously noted, we still expect any local star forming regions within an annulus to themselves be shielded), but are, on average, borderline unstable.  This behavior is similar to that observed in \citet{Orr17}, although we find lower values of $\qh$ in star forming regions.  This discrepancy is largely due to our different treatment of the stellar velocity dispersion. \citet{Orr17} assume $\sigma_*\sim c_s$, while we estimate values of $\sigma_*$ which are 2--3 times larger.

These data can also be used to test the behavior of $\qh\tau$.  $\qh\tau$ drops off rapidly with radius, reaching values $<$1 by ${\sim}R_e$ for all galaxies. These low values support the ``fragmentation-driven'' scenario discussed in Section~\ref{sec:methods}.  This interpretation is consistent with the behavior of $\qh$ and $\tau$ individually; as radius increases, $\tau$ continues to decline while $\qh$ remains $\sim1$, signaling that the outer disc would reach a point of instability well before it reached the threshold for self-shielding.

The $\tau$ parameter is not the only means of estimating where the disc is H$_2$ dominated.  Alternatively, \citet{Blitz06} provide a calibration relating $\Sigma_{\rm H_2}/\Sigma_{\rm \textsc{H\,i}}$ to hydrostatic midplane pressure, which is itself a function of gas and stellar surface densities and velocity dispersions.  We explored using this parameter instead of $\tau$, but find that the threshold where $\Sigma_{\rm H_2}/\Sigma_{\rm \textsc{H\,i}}=1$ does not change significantly.

\begin{figure*}
\includegraphics[width=2\columnwidth]{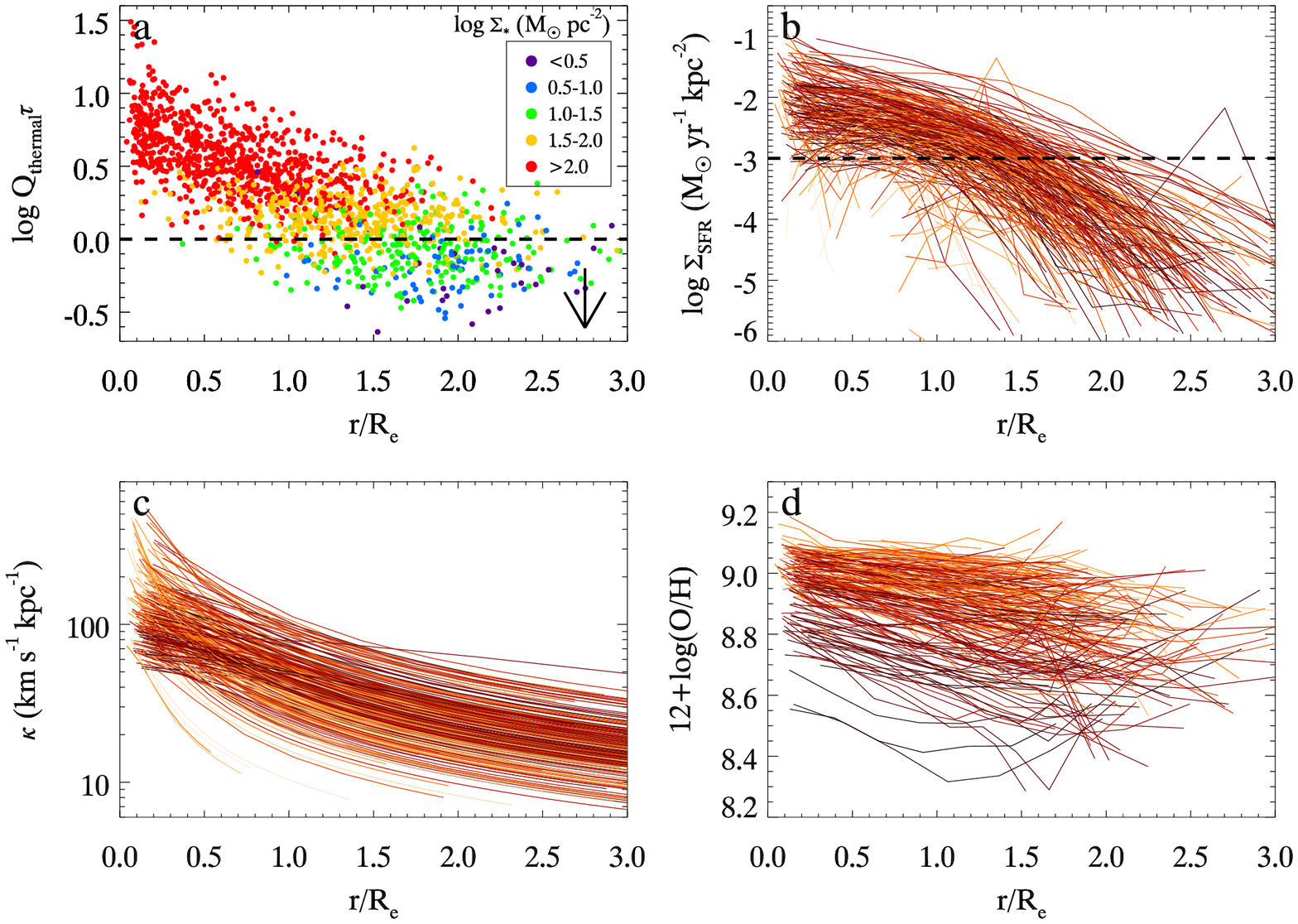}
\caption{(a) $\qh\tau$ vs. radius normalized by $R_e$, where $\qh\tau$ is approximated with Eq.~\ref{eq:qtau_approx}.  Each point represents one 2.5$\arcsec$ annulus where the N2O2 method is used to estimate $Z'$. The colours indicate $\Sigma_*$ in each annulus.  The arrow indicates the average amount that $\qh\tau$ would decrease if the O3N2 method was instead used to estimate $Z'$.  (b) Stacked $\Sigma_{\rm SFR}$ profiles for our MaNGA sample determined from H$\alpha$ surface brightness. (c) Stacked $\kappa$ profiles for our MaNGA sample determined from ionized gas rotation curves.  (d) Stacked $12+\log{\rm O/H}$ profiles determined with the N2O2 strong line method. For easier viewing, stacked profiles are colored based on their stellar mass, with lower mass galaxies shown in darker shades.}
\label{fig:tau_q_data}
\end{figure*} 

\subsection{The MaNGA Sample}

Although the L08 sample provides high quality data, it is extremely small and not representative of the full galaxy population. Therefore, we now examine the behavior of the much larger MaNGA sample. We cannot examine $\qh$ and $\tau$ directly with MaNGA, but we can constrain $\qh\tau$ using Eq.~\ref{eq:qtau_approx}. 

In Fig.~\ref{fig:tau_q_data}, we show the radial dependence of $\qh\tau$. Each small point represents one 2.5$\arcsec$ annulus in a single galaxy. We also show $\kappa$ and $12+\log{\rm O/H}$, the two quantities go directly into the calculation of $\qh\tau$, as well as $\Sigma_{\rm SFR}$, where $\Sigma_{\rm SFR}$ is estimated from H$\alpha$ luminosity using the calibration from \citet{Kennicutt12}.

$\qh\tau$ approaches values around $\sim$1 as we move away from the centres of galaxies, although there is significant scatter. In contrast to the analysis of the L08 sample, a significant number of galaxies have $\qh\tau>1$ at large radii. However, recall that unless $\Sigma_{\rm g} \gg \Sigma_*$, Eq.~\ref{eq:qtau_approx} is an upper limit on $\qh\tau$.  Not accounting for any stellar component can significantly overestimate $\qh\tau$, as illustrated by the profiles with and without the stellar component in Fig.~\ref{fig:qtau_wgas}.  We see evidence for this bias when we examine the behavior of the data in Fig.~\ref{fig:tau_q_data} as a function of $\Sigma_*$, where at fixed radius, annuli with lower $\Sigma_*$ tend to have lower $\qh\tau$. If we limit our analysis to annuli with $\Sigma_*<10^{1.5}\,M_{\odot}\,{\rm pc^{-2}}$, $\qh\tau<1$ on average.  Given that this subset yields more accurate estimates of $\qh\tau$ compared to the full sample when applying Eq.~\ref{eq:qtau_approx} (assuming $\Sigma_{\rm g}$ makes up a larger fraction of the disc surface densities in these regions), these data lend support to the ``fragmentation-driven" star formation scenario.  In Section{~\ref{sec:qtau_overest}, we revisit the issue of how much $\qh\tau$ may be overestimated in the presence of an unaccounted for stellar component.
 
An additional important systematic error to keep in mind is our choice of strong-line metallicity indicator. The R$_{23}$ and N2O2 methods we use for Fig.~\ref{fig:qtau_wgas} and \mbox{Fig.~\ref{fig:tau_q_data}} provide higher metallicities than most other strong-line calibrations.  For instance, in Fig.~\ref{fig:tau_q_data} we expect the upward systematic uncertainty on $\qh\tau$ to be $<0.1$ dex based on the comparison of different methods by \mbox{\citet{Kewley08}}. The larger uncertainty is downward, and the arrow in \mbox{Fig.~\ref{fig:tau_q_data}} illustrates the mean shift in the data points if we instead adopt the O3N2 strong-line metallicity calibration.  Additionally, our assumed gas sound speed is higher than typically used values for gravitational stability analysis. Although this is by design since we are interested in the transition from stable, unshielded, non-star forming $T=10^4\,K$ disc to one with the opposite properties, it also means $\qh$ makes optimistic assumptions about the ability of the gas to resist collapse. In summary, our combination of systematic errors make it more likely that we have overestimated, rather than underestimated, $\qh\tau$, which only strengthens our claims that $\qh\tau<1$ in the outer disks of galaxies.

\section{Discussion} \label{sec:discussion}

Using two independent data sets, we have explored the behavior of $\qh$, $\tau$, and $\qh\tau$ within galaxy discs in order to constrain whether they first reach the threshold for self-shielding or gravitational instability in star forming regions.  In the first data set from L08, we find star formation proceeding in annuli which are borderline unstable but far from self-shielded on average. With our second data set from the MaNGA survey, we find that the mean value on the {\it upper limit} of $\qh\tau$ at the edge of discs with widespread star formation is $\sim$1, but when focusing only on those galaxies with the lowest $\Sigma_*$ where the upper limits are most robust, we find $\qh\tau<1$ on average.  We begin our discussion of these results by first exploring the extent to which $\qh\tau$ may be overestimated by Eq.~\ref{eq:qtau_approx}, and how our results may change when we account for this effect.  We finish by discussing the implications our results have for what drives widespread star formation in galaxies.

\subsection{Overestimation of $\qh\tau$}
\label{sec:qtau_overest}

In Section~\ref{sec:results}, we discussed how $\qh\tau$ may be overestimated in the presence of an ignored stellar component. We now quantify the magnitude of this effect and its impact on our results. 

Eq.~\ref{eq:qtau_approx} yields the most accurate estimates of $\qh\tau$ when $\Sigma_{\rm g} \gg \Sigma_*$. Within the L08 sample, we find that such conditions are actually very rare; only a small fraction of galaxies have regions with $\Sigma_{\rm g}/\Sigma_* > 1$, and typically only at $r \gtrsim 2\,{\rm R_e}$.  A major caveat is that the L08 sample is not representative so we cannot explicitly assume these conclusions hold for all MaNGA galaxies.

However, using all available annuli in the L08 sample, we can get a sense for how $\Sigma_{\rm g}/\Sigma_*$ varies as a function of $\Sigma_*$. Fig.~\ref{fig:heracles_qtau_overest} plots the distribution of gas fraction  $f_g={\Sigma_{\rm g}}/{\left(\Sigma_{\rm g}+\Sigma_*\right)}$ as a function of $\Sigma_*$ from L08 (we use all 23 galaxies from L08 regardless of whether metallicity information is available).  There is a wide spread in $f_g$ as a function of $\Sigma_*$, but there is a clear upper envelope in the point distribution.  Only at very low values of $\Sigma_*$ ($\log \Sigma_* \lesssim 0.8\,M_{\odot}\,{\rm pc^{-2}}$) do gas surface densities begin to dominate over stellar surface densities.  Using of the upper envelope of $f_g$ in Fig.~\ref{fig:heracles_qtau_overest}, we estimate the maximum possible values of $\Sigma_g$ as a function of $\Sigma_*$, using $f_g$ values\footnote{These correction factors are judged by eye.} of  $0.8$, $0.7$, $0.4$, $0.2$, and $0.2$ for $\log \Sigma_*$ regimes of ${<}0.5$, $0.5$-$1.0$, $1.0$-$1.5$, $1.5$-$2.0$, and ${>}2.0\,M_{\odot}\,{\rm pc^{-2}}$.  These estimates are conservative towards instabilities in the sense that the illustrate the minimum relative impact of any stellar component on $\qh$ while maximizing the potential for self-shielding. 

\begin{figure}
\includegraphics[width=\columnwidth]{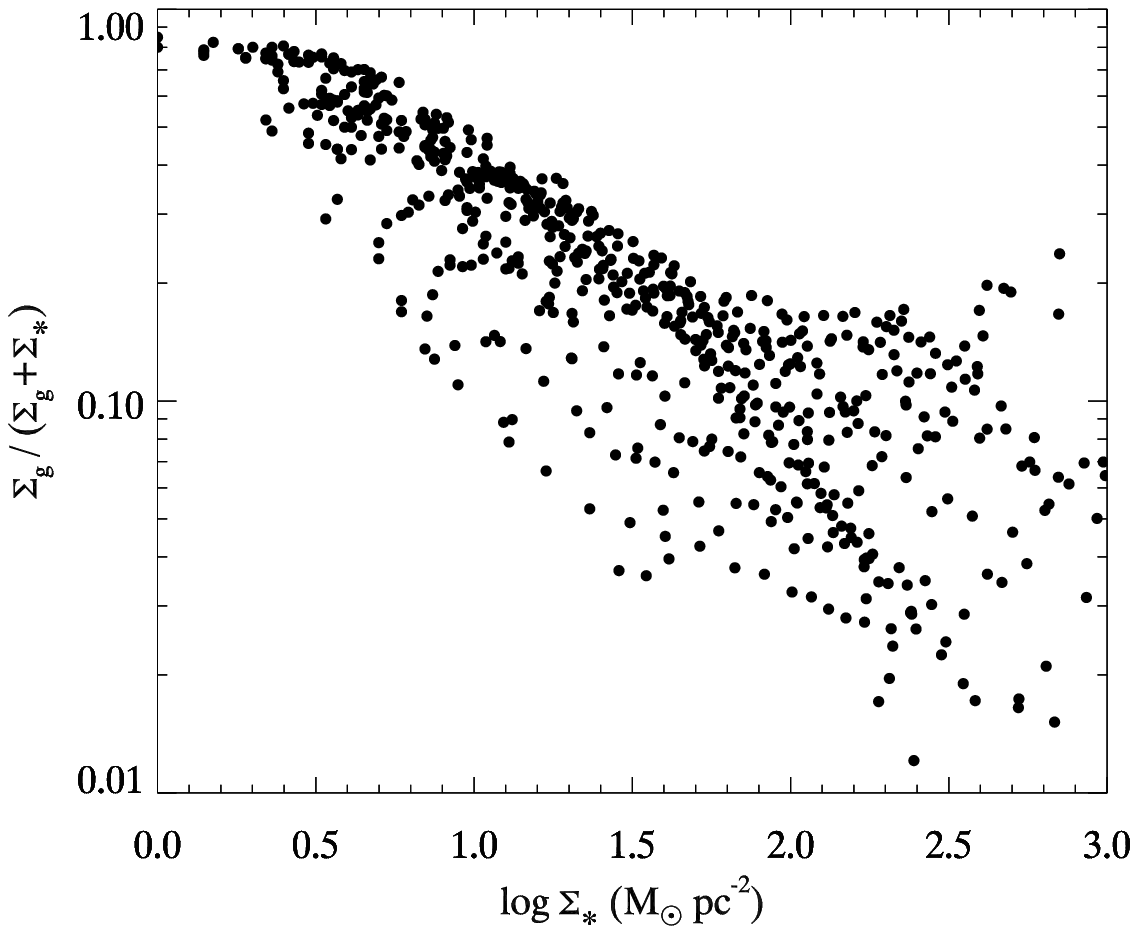}
\caption{The distribution of $\gamma=\left(1+\sfrac{\Sigma_*}{\Sigma_{\rm g}}\right)^{-1}$ as a function of $\Sigma_*$ for all annuli in the sample from \citet{Leroy08}.  This term determines the amount by which Eq.~\ref{eq:qtau_approx} should be multiplied by in order to obtain the true value of $\qh\tau$.}
\label{fig:heracles_qtau_overest}
\end{figure}

\begin{figure*}
\includegraphics[width=2\columnwidth]{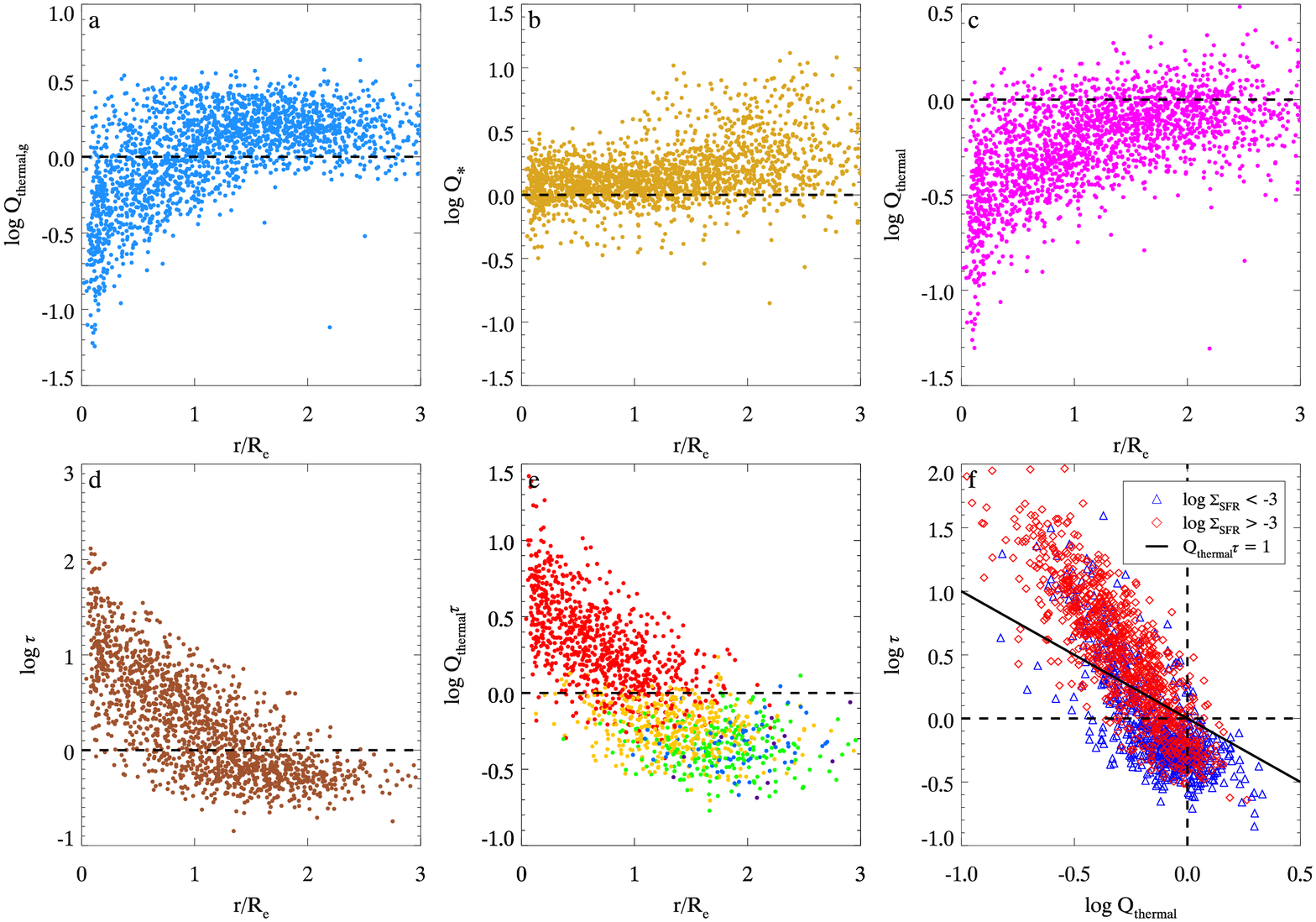}
\caption{Properties of our MaNGA sample after approximating $\Sigma_{\rm g}$ based on the data in Fig.~\ref{fig:heracles_qtau_overest} and incorporating the stellar component.  Panels (a)--(e) show the radial dependence of $\qgh$,  $Q_*$, $\qh$, $\tau$, and $\qh\tau$. The color-coding in panel (e) is the same as that of Fig.~\ref{fig:tau_q_data}a. Panel (f) shows the distribution of annuli in $\tau$,$\qh$ parameter space. This analysis supports our claim that disc will reach the threshold for disc stability before they reach the threshold for self-shielding.}
\label{fig:manga_qtau_corr}
\end{figure*}

Fig.~\ref{fig:manga_qtau_corr} shows radial profiles of $\qgh$, $Q_*$, $\qh$, $\tau$, and $\qh\tau$ using our estimates of $\Sigma_{\rm g}$.  As expected, the combination of gas and stars has lowered $\qh$ such that it largely falls below unity at large radius. Similarly, $\tau<1$ at large radius, even though our estimates of $\Sigma_g$ are the maximum possible values. Additionally, $\qh\tau<1$ consistently in the outer disc, and the secondary dependence on $\Sigma_*$ is much weaker compared to Fig.~\ref{fig:tau_q_data}.  In summary, this analysis supports the fragmentation-driven scenario by showing that annuli will cross the threshold for gravitational instability before they cross the threshold for self-shielding.

Fig.~\ref{fig:manga_qtau_corr} also shows the distribution of annuli in $\tau$ vs. $\qh$ parameter space, which is the observational equivalent to Fig.~8 from \citet{Orr17} who examine this distribution in the FIRE simulations. Consistent with \citet{Orr17}, we find that star forming annuli \mbox{($\Sigma_{\rm SFR}>10^{-3}\,M_{\odot}\,{\rm yr^{-1}\,kpc^{-2}}$)} begin to appear around $\qh\sim1$ and $\tau<1$. Our data do not show the locus on non-star forming annuli extending into the $\qh>1$,$\tau>1$ quadrant shown by \citet{Orr17}, but this is due observational limitations (i.e., we cannot measure rotation velocity or metallicity without the presence of emission lines). Independent of how we divide star-forming and non-star forming annuli, this figure illustrates how the regime with $\tau>1$ and $\qh>1$ is essentially unpopulated, emphasizing the point that galaxies appear to first reach gravitational instability and then self-shield, not vice versa.

The above analysis has been done assuming the highest reasonable value of $\Sigma_{\rm g}$ at a given $\Sigma_*$.  Although this choice maximizes $\tau$, it biases us towards lower $\qh$. However, $Q_*$ sets an upper limit on $\qh$, and we estimate $Q_*$ to be only slightly above 1.  Therefore, any decrease in $\Sigma_g$ will lead to an increase in $\qh$ that is much slower than the corresponding decrease in $\tau$.  Therefore, even with lower estimates for $\Sigma_{\rm g}$, our discs will still be much closer to the threshold for gravitational instability than the threshold for self-shielding.

\subsection{Implications for Star Formation in Galaxy discs}

Through both direct means (via measurements of $\qh$ and $\tau$) and indirect means (using $\qh\tau$), we find results consistent with the existence of widespread star formation in galaxy discs in regions that are on-average not self-shielded, but are unstable (or at least borderline unstable) in the presence of thermal gas support.  Regardless of whether or not the discs are formally gravitationally unstable, their distance from the threshold for disc instability is far smaller than their distance from the threshold for self-shielding, indicating that an annulus will first become unstable and fragment before it becomes self-shielded.  These data are consistent with  a star formation scenario where widespread star formation in galaxy discs begins once a disc can no longer thermally support itself, at which point it fragments and triggers the growth of dense clouds which can locally self-shield.  Our data are in agreement with \citet{Orr17} that examines the onset of star formation within the FIRE galaxy simulations.  

Our results may appear somewhat at odds with the findings of \citet{Schaye04} who argue that it is the formation of a molecular phase (even at very low levels) that leads to disc instabilities due to the associated sudden drop in gas temperature. Our findings essentially argue that such a drop in temperature is not a necessary ingredient to obtain disc instabilities; even when the gas is assumed to be $10^4$K, the discs are at least borderline unstable. Similar results are seen in \citet{Orr17} where they find widespread gravitational collapse and star formation even when gas cooling is turned off. It is important to note that \citet{Schaye04} associate this cooling with $\Sigma_{\rm H_2}/\Sigma_{HI}\ll1$ and do not argue that the disc is fully self-shielded by when the cooling takes place, and in their models, $\Sigma_{\rm H_2}/\Sigma_{HI}=1$ occurs at a much smaller radius.  Therefore, our results are not entirely inconsistent with the findings of \citet{Schaye04}, although we do not require any explicit drop in gas temperature/$c_s$ to lead to disc instabilities.  If such temperature drops do occur, this will lead to lower $Q$ and strengthen our results.

Although this study provides support for gravitational instabilities as the initial trigger that eventually leads to star formation in galaxy discs, we are not making any claims about the local efficiency of star formation once it begins, nor are we dismissing the importance of self-shielding on local scales.  We are simply examining the large-scale conditions within galaxies that are conducive to dense cloud formation and widespread star formation.  We claim that $\qh$ is the more fundamental condition for star formation on large scales, and this is because the star-forming disc appears to become on-average unstable before it becomes on-average self-shielded. None the less, even in annuli which are on average not self-shielded, self-shielding is undoubtedly instrumental in local star forming regions.  

Our study does not explicitly explain the very inefficient star formation that occurs in extreme outer discs \citep{Bigiel10}.  Based on the radial profiles of $\qh$ in Fig.~\ref{fig:qtau_wgas}, we expect the extended regions of galaxies to have $\qh>1$, consistent with other studies which find outer discs to generally be dynamically stable \citep{Kennicutt89, Martin01}.  As discussed in Section~\ref{sec:intro} it is possible for star formation to occur locally in a disc that is on-average stable and unshielded if localized overdensities can be created. Without explicitly arguing what may drive such overdensities, we do note that we measure values of $\qh\tau$ that continue to decrease below unity at large radius.  Even though these discs are neither gravitationally unstable or self-shielded, $\qh\tau<1$ implies that an increase in gas density will lead to gravitational instability before self-shielding, so the fragmentation-driven picture of star formation may still apply in the extreme outer disks of galaxies.

\section{Conclusions} \label{sec:conclusions}
We have presented a new analysis on the relative importance of gravitational disc instabilities (parametrized by $\qh$) and self-shielding (parametrized by $\tau$) to the onset of star formation in galaxy discs.  We consider two basic scenarios for star formation where the primary condition for widespread star formation differs: the ``self shielding-driven'' model where discs first self-shield themselves against the background UV field and then become unstable and fragment, and the ``fragmentation-driven'' model where discs first become unstable and fragment, only after which they are able to self-shield. 

Using a small sample of galaxies with high-quality gas data where we can independently examine $\qh$ and $\tau$, we find evidence that galaxies will be able to reach the threshold for gravitational instability well before they cross the threshold for self-shielding in their outer discs.  Using a larger sample from the MaNGA survey lacking direct gas information, we show that the value of $\qh\tau$ (which can be constrained in the absence of direct gas measurements) is consistent with galaxies being able to cross the threshold for disc instabilities first.  The results from both of these samples lend support to the ``fragmentation-driven'' scenario for star formation in galaxies.

Future observations that can directly resolve gas distributions in large and diverse samples of galaxies out to large radii will be highly valuable to further test the fragmentation-driven star formation scenario.  Wide field interferometric surveys like WALLABY and APERTIF which overlap MaNGA and other IFU surveys will be particularly beneficial. Similarly, the upcoming Local Volume Mapper (LVM) will yield high quality optical spectroscopy to compliment the growing radio/mm-wave inventory of nearby galaxies. The combination of complete data for large samples will also enable exploration into whether the behavior of $\qh$, $\tau$, and $\qh\tau$ are universal for all galaxies, or whether they vary with other galaxy properties or their environments, potentially providing additional insight into the physical processes that regulate widespread star formation in galaxy discs.

\section*{Acknowledgements}
We thank our anonymous referee for their constructive feedback which greatly improved this work. We would also like to
thank Christy Tremonti and Eric Emsellem for useful discussions. This
work was supported by World Premier International Research centre
Initiative (WPI Initiative), MEXT, Japan. MAB acknowledges NSF-AST-1517006. AW acknowledges
support of a Leverhulme Trust Early Career Fellowship. DB acknowledges support from RSF grant RSCF-14-50-00043. MEO was supported by the National Science Foundation Graduate Research Fellowship under Grant No. 1144469. Funding for the
Sloan Digital Sky Survey IV has been provided by the Alfred P. Sloan
Foundation, the U.S. Department of Energy Office of Science, and the
Participating Institutions. SDSS-IV acknowledges support and resources
from the centre for High-Performance Computing at the University of
Utah. The SDSS web site is www.sdss.org.

SDSS-IV is managed by the Astrophysical Research Consortium for the 
Participating Institutions of the SDSS Collaboration including the 
Brazilian Participation Group, the Carnegie Institution for Science, 
Carnegie Mellon University, the Chilean Participation Group, the French Participation Group, Harvard-Smithsonian centre for Astrophysics, 
Instituto de Astrof\'isica de Canarias, The Johns Hopkins University, 
Kavli Institute for the Physics and Mathematics of the Universe (IPMU) / 
University of Tokyo, Lawrence Berkeley National Laboratory, 
Leibniz Institut f\"ur Astrophysik Potsdam (AIP),  
Max-Planck-Institut f\"ur Astronomie (MPIA Heidelberg), 
Max-Planck-Institut f\"ur Astrophysik (MPA Garching), 
Max-Planck-Institut f\"ur Extraterrestrische Physik (MPE), 
National Astronomical Observatories of China, New Mexico State University, 
New York University, University of Notre Dame, 
Observat\'ario Nacional / MCTI, The Ohio State University, 
Pennsylvania State University, Shanghai Astronomical Observatory, 
United Kingdom Participation Group,
Universidad Nacional Aut\'onoma de M\'exico, University of Arizona, 
University of Colorado Boulder, University of Oxford, University of Portsmouth, 
University of Utah, University of Virginia, University of Washington, University of Wisconsin, 
Vanderbilt University, and Yale University.



\bibliographystyle{mnras}
\bibliography{mybib}








\bsp	
\label{lastpage}
\end{document}